\documentclass[trackchanges,twocolumn]{aastex701}

\usepackage{amsmath}
\usepackage{hyperref}
\usepackage{media9}

\begin{document}

\title{HESS J1507-622: A Plausible Young Galactic Kilonova Remnant}

\author[orcid=0000-0003-3159-7148]{Prantik Sarmah}
\affiliation{State Key Laboratory of Particle Astrophysics, Institute of High Energy Physics, Chinese Academy of Sciences, Beijing, 100049, China}
\email{prantiksarmah@ihep.ac.cn}  

\author[orcid=0000-0002-5901-98792]{Xilu Wang} 
\affiliation{State Key Laboratory of Particle Astrophysics, Institute of High Energy Physics, Chinese Academy of Sciences, Beijing, 100049, China}
\email{wangxl@ihep.ac.cn}

\author[orcid=0009-0009-2801-6374]{Myles Carley}
\affiliation{Department of Physics and Astronomy, University of Notre Dame, Notre Dame, IN 46556, USA}
\email{mcarley@nd.edu}

\author[0000-0001-7599-0174]{Shu-Xu Yi} 
\affiliation{State Key Laboratory of Particle Astrophysics, Institute of High Energy Physics, Chinese Academy of Sciences, Beijing, 100049, China}
\email{sxyi@ihep.ac.cn}

\author[orcid=0000-0002-3776-4536]{Ming-Yu Ge} 
\affiliation{State Key Laboratory of Particle Astrophysics, Institute of High Energy Physics, Chinese Academy of Sciences, Beijing, 100049, China}
\email{gemy@ihep.ac.cn}

\author[0000-0002-4729-8823]{Rebecca Surman}
\affiliation{Department of Physics and Astronomy, University of Notre Dame, Notre Dame, IN 46556, USA}
\email{rsurman@nd.edu}

\correspondingauthor{Prantik Sarmah, Xilu Wang}
\email{prantiksarmah@ihep.ac.cn, wangxl@ihep.ac.cn}

\begin{abstract}
HESS J1507$-$622 is an unusual TeV gamma-ray source in the H.E.S.S. Galactic Plane Survey, distinguished by its off-plane location ($b \simeq -3.5^{\circ}$) and compact angular extent ($\sim 0.3^{\circ}$). Its location and morphology challenge conventional pulsar wind nebula and supernova remnant interpretations, and the absence of a detected shell or associated pulsar leaves its origin unresolved. Here, we propose that HESS J1507$-$622 is a young Galactic kilonova remnant with gamma-ray emission of leptonic origin. In this scenario, the source's off-plane location is naturally explained by the natal kick imparted to the progenitor binary neutron star system. We demonstrate that a leptonic kilonova remnant model can successfully reproduce the observed GeV-TeV flux via inverse-Compton scattering from shock-accelerated electrons, while the corresponding synchrotron emission remains below current X-ray and radio upper limits. Based on the source's energetics and angular size, we infer a distance of 3.8-14.3 kpc and an age of 0.2-3.0 kyr. Importantly, our analysis favors a kilonova remnant younger than 1 kyr. The lack of a historical naked-eye record can be explained by the transient's rapid fading, far-southern declination, and potential line-of-sight extinction. We further assess the detectability of unique multi-wavelength signatures that could confirm this hypothesis. We find promising prospects for detecting the unique MeV gamma rays from the decay of $r$-process radioisotopes using future observatories, alongside complementary low-energy signatures from thermal dust emission and a possible optical light echo. These distinct signals offer definitive tests of the kilonova remnant interpretation and motivate targeted observations with current and future facilities.
\end{abstract}


\section{Introduction}
Kilonovae (KNe), often  associated with short gamma-ray bursts (SGRBs), are transient events occurring during the merger of compact binaries, namely binary neutron stars or black hole-neutron star systems~\citep{2014ARA&A..52...43B,2007PhR...442..166N}. 
\footnote{However, the connection between merger progenitors and observed burst properties is not always straightforward: the observed duration is shaped by a variety of physical factors beyond the central engine activity~\citep{2025JHEAp..45..325Z}, a distinct subclass of GRBs originating from compact binary mergers challenges the traditional duration-based classification~\citep{2025ApJ...979...73W}, and a short-lived central engine can power long-duration prompt emission under certain conditions~\citep{2025JHEAp..4700359Y,2026JHEAp..5300607Y}.} 
KNe are now recognized as a primary site of rapid neutron-capture ($r$-process) nucleosynthesis, producing a substantial fraction of the universe's elements heavier than iron~\citep{Pian:2017gtc,Savchenko:2017ffs,Cote+2018}.
Motivated by the GW170817/GRB 170817A observations, recent studies~\citep{Kimura:2018ggg,Rodrigues:2018bjg} suggest that KNe may contribute non-negligibly to cosmic rays (CRs) at very high energies. Such CRs are expected to generate secondary high-energy gamma rays and neutrinos  through hadronic and photo-hadronic interactions during their propagation to Earth. However, because the KN rate is much lower than the core-collapse supernova (SN) rate~\citep{Eldridge:2018nop}, their contribution to the diffuse gamma-ray and high-energy neutrino backgrounds is expected to be subdominant, which limits our understanding of CR interactions in these environments. 
Therefore, these interactions can be probed only for nearby point sources, i.e, either a newborn KN or the remnant of a past event.  
Given a Galactic KN rate of $(10^{-5}-10^{-4})$~$\rm yr^{-1}$~\citep{Kim:2006fm,LIGOScientific:2018mvr}, only a few KN remnants (KNRs) are expected in the Milky Way, compared to hundreds of SN remnants (SNRs)~\citep{Green:2024uci,Mantovanini_2025,Ranasinghe:2022ntj}. To date, no such Galactic KNR has been detected. 

The High Energy Stereoscopic System (H.E.S.S.) in Namibia has detected numerous very-high-energy gamma-ray sources (VHE)~\citep{HESS:2018pbp}, several of which remain unidentified. While most lie along the Galactic plane, HESS J1507$-$622 ($l = 317.97^{\circ}$, $b = -3.47^{\circ}$) is a notable off-plane source. Its origin remains debated: interpretation as an old ($>10$ kyr) pulsar wind nebula (PWN)~\citep{2011A&A...525A..45H,Domainko2011AdSpR..47..640D,2012A&A...545A..94D} is hindered by the lack of an associated pulsar, while a SNR origin  
is challenged by the non-detection of a shell~\citep{2012A&A...545A..94D}. Furthermore, HESS J1507$-$622 lacks confirmed low-energy counterparts in X-ray, radio, or optical bands. Although this absence of synchrotron emission might simply be due to a weak ambient magnetic field, the cumulative limitations of standard PWN and SNR models invite alternative explanations.

As first suggested but not explored by \citet{2011A&A...525A..45H}, this source could be a Galactic KNR. A binary neutron star (BNS) merger naturally explains the off-plane location: assuming a typical natal  kick velocity of $\sim (10-100)~\rm km/s$~\citep{Wong2010}, a BNS originating in the Galactic bulge would take about $0.1-1$~Gyr to reach the location of J1507$-$622, consistent with the inferred  typical BNS merger delay times~\citep{Beniamini:2019iop}.

In this Letter, we present a detailed investigation of the potential KN origin of HESS J1507$-$622. Based on its observed properties, we develop a KNR model and constrain the relevant physical parameters. We demonstrate that a leptonic emission scenario within this framework can successfully reproduce the observed VHE gamma-ray flux. Furthermore, we show that late-time MeV gamma-ray emission, powered by the decays of radioactive $r$-process nuclei, could be detectable by next generation missions such as the MeV Gamma-Ray Observatory (MeVGRO), providing a definitive test of the KN origin. We also highlight promising prospects for detecting multi-wavelength emission associated with KN dust. Overall, this study of HESS J1507$-$622 points out the potential of coordinated multi-messenger and multi-wavelength observations to unveil the nature of this intriguing source.

The manuscript is organized as follows: Section~\ref{sec:HESS_source} reviews the observed properties of HESS J1507$-$622 and the limitations of previous interpretations. Section~\ref{sec:KN_Model} details our KN model and constraints, followed by an exploration of multi-wavelength detectability in Section~\ref{sec:KN_detectability}. Finally, we conclude in Section~\ref{sec:conclusion}.

\section{HESS J1507$-$622}
\label{sec:HESS_source}
\begin{figure*}
    \centering
    \includegraphics[width=0.9\linewidth]{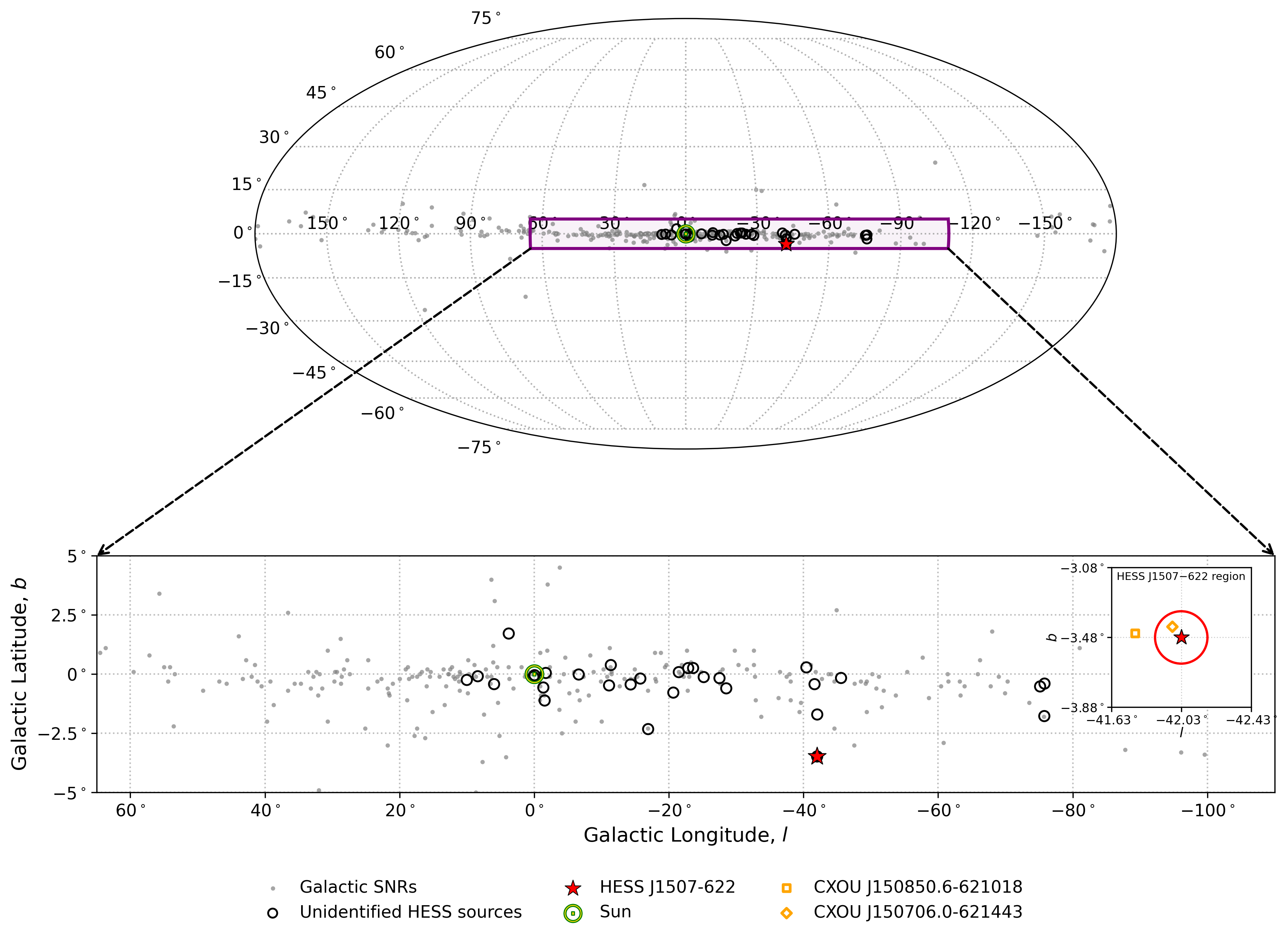}  
    \caption{The top panel shows the unidentified sources from the HGPS catalog (black circles). The bottom panel shows the zoomed in region shown by the purple box in the top panel. The source HESS J1507-622 is represented by the red star in both panel. The grey points show 310 Galactic SNRs from \cite{Green:2024uci}. The inset plot shows the  region of the HESS source (red circle) together with the two X-ray sources in this region CXOU J150850.6$-$621018 (orange square) and CXOU J150706.0$-$621443 (orange diamond).}
    \label{fig:HESS_sources}
\end{figure*}

The 2004-2013 H.E.S.S. Galactic Plane Survey~\citep{HESS:2018pbp} discovered 78 very-high-energy (VHE) gamma-ray sources, 30 of which remain unclassified in the \texttt{VizieR}\footnote{{https://astroquery.readthedocs.io/en/latest/vizier/vizier.html}} catalog (using python interface ``Astroquery"). In Fig.~\ref{fig:HESS_sources}, we plot these 30 unidentified sources (black circles) for comparison with 310 cataloged Galactic SNRs~\citep{Green:2024uci} (gray dots). We highlight source J1507$-$622 with a red star and discuss its observational properties and implications below.

\subsection{Observational properties of HESS J1507$-$622}

HESS J1507$-$622 is a steady, unidentified, extended VHE gamma-ray source at the Galactic coordinates $(l,b)\simeq(317.97^\circ,-3.49^\circ)$ (centroid ${\rm RA}=15^{\rm h}06^{\rm m}53^{\rm s}$ and ${\rm Dec}=-62^\circ20'59''$, J2000). Its $b \simeq -3.5^\circ$ off-plane displacement is considerably larger than that of the majority of H.E.S.S. Galactic VHE sources. Detected by H.E.S.S. at $>9\sigma$, it has an intrinsic Gaussian width of $\simeq0.15^\circ$ and lacks significant VHE variability, favoring an extended Galactic source over a variable compact object \citep{2011A&A...525A..45H}.

Its VHE gamma-ray spectrum ($>1$~TeV) follows a power law $dN_\gamma/dE\propto E^{-\Gamma}$ with $\Gamma=2.24\pm0.16_{\rm stat}\pm0.20_{\rm sys}$ and integral flux $(1.5\pm0.4_{\rm stat}\pm0.3_{\rm sys})\times10^{-12}~{\rm cm^{-2}s^{-1}}$ \citep{2011A&A...525A..45H}. Using $\sim34$ months of \textit{Fermi}-LAT data, the 0.3-300~GeV emission smoothly connects to the TeV spectrum with a hard index $\Gamma=1.7\pm0.1_{\rm stat}\pm0.2_{\rm sys}$ and integrated flux $(2.0\pm0.5_{\rm stat}\pm1.0_{\rm sys})\times10^{-9}~{\rm cm^{-2}s^{-1}}$ \citep{2012A&A...545A..94D}. 

Despite its prominent GeV-TeV emission, HESS J1507$-$622 lacks a compelling lower-energy counterpart. Archival ROSAT data revealed only three faint X-ray sources near or just outside the TeV boundary \citep{2011A&A...525A..45H}. Dedicated observations of \textit{Chandra}, \textit{XMM-Newton}, and \textit{Suzaku} identified an extended source, CXOU J150850.6$-$621018, located $\simeq20'$ from the TeV centroid (shown as orange square in Fig.~\ref{fig:HESS_sources}); however, this substantial spatial offset makes a direct association problematic \citep{2011A&A...525A..45H,Eger:2014eaa}. While another closer source CXOU J150706.0$-$621443 (shown as orange diamond in Fig.~\ref{fig:HESS_sources}) exhibits a hard X-ray spectrum, limited photon counts prevent a firm association \citep{2014A&A...567A..74T, Eger:2014eaa}. Crucially, no diffuse X-ray emission matching the TeV morphology has been found \citep{2014A&A...567A..74T, Eger:2014eaa}.

Similarly, Molonglo Galactic Plane Survey data reveal no spatially coincident radio source or shell\citep{2011A&A...525A..45H}. Furthermore, a large-scale 2.4~GHz radio filament and a CO cloud located in the region are both too extended and too far offset from the TeV centroid to suggest a physical association \citep{2011A&A...525A..45H}. 

Unlike most firmly identified gamma-ray SNRs, HESS J1507$-$622 lacks any independently detected shell structure in radio or X-ray. Its observational profile is therefore highly unusual: it features bright, extended ($0.3^\circ$) GeV-TeV emission but extremely faint or absent lower-energy counterparts. This exceptionally large gamma-ray to X-ray flux ratio, combined with its location off the Galactic plane, distinguishes it as one of the most mysterious unidentified Galactic VHE sources. Consequently, previous studies have proposed various interpretations for its origin, including an evolved PWN, an SNR, and more exotic transient populations \citep{2011A&A...525A..45H,2012A&A...545A..94D,2014A&A...567A..74T,Eger:2014eaa}.

\subsection{Pulsar Wind Nebulae Explanation and Limitations}
\label{subsec:PWN_explanation}

Based on its observed gamma-ray spectrum, HESS J1507-622 has often been interpreted as an evolved pulsar wind nebula (PWN). However, this scenario faces several challenges, which we briefly discuss below.

\textit{Limited energy budget--}
While a low local interstellar medium (ISM) density disfavors a hadronic origin, a leptonic inverse-Compton (IC) scenario is plausible but energetically constrained. 
Initial pulsar spin periods of $P_0=10$--$100$~ms yield rotational energies of $E_{\rm rot}\sim2\times10^{48}$--$2\times10^{50}$~erg as $E_{\rm rot} \approx 2\times10^{50}\,(P_0/10~\mathrm{ms})^{-2}~\rm erg$~\citep{2012A&A...545L...2H}. However, the total required electron energy, $E_e\approx3\times10^{47}(d/1~\mathrm{kpc})^2$~erg~\citep{2012A&A...545A..94D}, combined with an acceleration efficiency of $\lesssim10\%$~\citep{Zhang:2020kgg}, requires $E_{\rm rot}\gtrsim3\times10^{48}(d/1~\mathrm{kpc})^2$~erg. At a distance of $d=5$~kpc, this corresponds to $E_{\rm rot}\gtrsim7.5\times10^{49}$~erg with $P_0\lesssim16$~ms, demanding a typical pulsar birth period. For distances $d\gtrsim10$~kpc, the required energy exceeds $2\times10^{50}$~erg, making the PWN interpretation energetically untenable. 

\textit{Constraints from correlation of spin-down power, pulsar age, TeV luminosity, and  off-plane location--}
Another major limitation of the PWN hypothesis is the non-detection of any associated pulsar  within the expected spatial region around the HESS source~\citep{2012A&A...545A..94D,HESS:2017lee}. Furthermore, reproducing the observed gamma-ray flux requires a PWN age of $\sim 10^5$ yrs \citep{2012A&A...545A..94D}. This pulsar age implies that the required spin-down power ($\dot{E}$) corresponds to $\sim (10^{33}-10^{36})~\rm erg/s$~\citep{HESS:2017lee}. 
Then the empirical correlation between TeV luminosity ($L_{\rm 1-10 ~\rm TeV}$) and the spin-down power reported in~\cite{HESS:2017lee} yields a TeV luminosity of $\sim  (3 \times 10^{31} -2 \times 10^{33})~\rm erg/s$. This together with the observed flux of $\sim 2 \times 10^{-9}~\rm GeV\,cm^{-2}\,s^{-1}$~\citep{2011A&A...525A..45H} in the energy range $(1$--$10)$~TeV infers a relatively close distance of $\sim (0.3-2.0)$~kpc. A source this close would locate near the Galactic plane, where the ambient magnetic field is expected to be several $\mu\rm G$~\citep{2012A&A...545A..94D,Eger:2014eaa,2011A&A...525A..45H}. The resulting strong synchrotron emission is in tension with the non-detection of any X-ray or radio counterparts for HESS J1507$-$622.

\subsection{Supernova Remnant Explanation and Limitations}
In addition to a PWN scenario, a Galactic SNR represents another possible origin for HESS J1507$-$622~\citep{Domainko2011AdSpR..47..640D}. However, unlike pulsars, SNRs do not generally acquire substantial natal kicks and therefore remain close to their progenitor explosion sites. 
While several SNRs have been detected at relatively high Galactic latitudes~\citep{Green:2024uci}, these are predominantly nearby systems (within a few kpc) where large angular offsets do not necessarily correspond to comparably large physical distances from the Galactic plane.
Therefore, if HESS J1507$-$622 were a SNR, its off-plane location would be particularly challenging to explain.
Additionally, the absence of any identified SNR shell in the radio or X-ray bands leaves the SNR association without independent observational support.

Given the limitations for both the PWN and SNR explanations, the origin of HESS J1507$-$622 remains an open question.

\section{Kilonova Remnant Explanation for HESS J1507$-$622}
\label{sec:KN_Model}

\citet{2011A&A...525A..45H} briefly considered a KN origin for HESS J1507$-$622 but disfavored it because a strictly hadronic origin for the observed gamma-ray emission would require an unrealistically high explosion energy. However, a leptonic scenario can produce these gamma rays with only a moderate explosion energy budget. Here, we demonstrate that a KN origin is indeed viable for this source by showing how a leptonic scenario can successfully reproduce the observed TeV gamma-rays and provide estimates for the unobserved lower-energy emission.  

\subsection{Cosmic-ray acceleration in Kilonova ejecta}
\label{subsec:CR_model}

Similar to the mechanism in SNRs, CRs can be accelerated to high energies in a KNR via diffusive shock acceleration as the expanding ejecta interacts with the ISM. While CR protons and heavy nuclei might reach PeV energies in this environment, they are unable to account for the observed gamma-ray flux due to the extremely low ambient ISM density at the off-plane location of HESS~J1507$-$622. The characteristic interaction length for CR protons is $\lambda = 1/n_{\rm ISM  } \sigma_{\rm pp} \approx 11$~Gpc, assuming the ISM number density $n_{\rm ISM} \sim 10^{-3}~{\rm cm^{-3}}$ and an inelastic $pp$ cross section of $\sigma_{\rm pp} \approx 3 \times 10^{-26}~{\rm cm^2}$. This huge interaction length implies that CRs escape the source long before undergoing hadronic interactions. Consequently, the hadronic gamma-ray and high-energy neutrinos are expected to be negligible.

In the leptonic scenario, the observed gamma-ray emission is dominated by IC scattering and synchrotron radiation from shock accelerated non-thermal electrons; bremsstrahlung is heavily suppressed by the low ISM density. The IC flux depends on the accelerated electrons' spectrum and target photons from the cosmic microwave background (CMB)~\citep{Fixsen:1996nj}, as the interstellar radiation field (ISRF) is negligible due to the off-plane location of HESS~J1507$-$622.
In contrast, the synchrotron emission is governed by the ambient magnetic field. While a standard exponential  Galactic magnetic field model~\citep{2024ApJ...966..240X} yields the magnetic field $B \sim (0.4\text{--}5)\,\mu{\rm G}$ at a distance of (1-20)~kpc, previous PWN models for this source~\citep{2012A&A...545A..94D,Eger:2014eaa,2011A&A...525A..45H} adopted $B\sim (0.5-1)~\mu\rm G$ to explain the weak lower-energy emission. Therefore, we consider this range in $B$ to predict the radio and X-ray fluxes. Additionally, because synchrotron emission may be substantially attenuated by unknown dust extinction, we neglect extinction and regard our predicted synchrotron fluxes as upper limits.

To estimate the gamma-ray flux, we assume the following spectral distribution of accelerated primary electrons,
\begin{equation}
\frac{d\phi_e}{dE_e}
=
\phi_{e,0}
\left(\frac{E_e}{E_{0}}\right)^{
-\left[
\alpha
+
\beta \log_{10}\!\left({E_e}/{E_{0}}\right)
\right]
}\,,
\label{eq:electron_spectrum}
\end{equation}
where $E_0\sim1$ TeV is the reference energy. The normalization constant $\phi_{e,0}$ is obtained from $\int_{\rm m_{e}}^{E_{e,\rm max}} E_e (d\phi_e/dE_e)  \mathrm{d}E_e = E_{e,\rm tot}$. The indices $\alpha$ and $\beta$ determine the spectral shape and are obtained by fitting the observed gamma-ray flux. We assume the total electron energy $E_{e, \rm tot}$ is a fraction $\varepsilon_{\rm e}$ of the total KN ejecta energy $E_{\rm ej}$. Note that in the PWN scenario, $E_{e, \rm tot}$ is determined from the total rotational energy of the pulsar~\citep{2012A&A...545A..94D}, as mentioned in Sec.~\ref{subsec:PWN_explanation}. 

The maximum electron energy, $E_{e, \rm max}$, is obtained by equating the acceleration timescale  with various loss timescales such as IC and synchrotron~\citep{1999ApJS..120..299T,1983RPPh...46..973D,Petropoulou:2016zar}. 
For a typical Sedov-Taylor phase KNR with $E_{\rm ej} \sim 10^{50}$~erg, the maximum electron  energies corresponding to IC and synchrotron losses are $E_{e, \rm max}^{\rm IC} \approx 244$~TeV~$v_{\rm sh}\,B^{1/2}$ and $E_{e, \rm max}^{\rm syn} \approx 769$~TeV~$v_{\rm sh}\,B^{-1/2}$, respectively. Here, $v_{\rm sh}$ is the shock velocity  in the unit of $10^4~\rm km\,s^{-1}$ and $B$ is in $\mu$G. 
For a typical KN model with $E_{\rm ej} \sim 10^{50}$~erg, $n_{\rm ISM} \sim 10^{-3}~\rm cm^{-3}$, and $t_{\rm age} \sim 1$~kyr; the shock velocity scales as $v_{\rm sh} \approx 10^4~\rm km\,s^{-1}$~$E_{\rm ej}^{1/5} \, n_{\rm ISM}^{-1/5} \, t_{\rm age}^{-3/5}$. 
Since $E_{\rm e, max}^{\rm IC} < E_{\rm e, max}^{\rm syn}$, the accelerated electron spectrum is governed by IC ($E_{\rm e, max}=E_{\rm e, max}^{\rm IC}$). We compute the resulting non-thermal emission from IC scattering and synchrotron radiation with the \texttt{GAMERA} code~\citep{2009arXiv0912.4229T}. 

\subsection{Kilonova parameter space}

The KN ejecta kinetic energy, $E_{\rm ej} = M_{\rm ej} V_{\rm ej}^2/2$, is defined by the ejecta mass $M_{\rm ej}$ and velocity $V_{\rm ej}$. 
As the ejecta expands into the surrounding ambient medium, it gives rise to strong shocks capable of accelerating charged particles, including electrons, protons, and heavier nuclei. Consequently, the efficiency of particle acceleration is strongly influenced by the properties of the ejecta. Additionally, the angular size of HESS J1507$-$622 is directly related to both the physical radius of the TeV-emitting region and the age of the KN remnant. Thus, to understand whether HESS J1507$-$622 could originate from a past KN event, we must determine if any physical $M_{\rm ej}-v_{\rm ej}$ combination can simultaneously reproduce the observed TeV gamma-ray emission and the angular size of HESS J1507$-$622 ($\theta_{\rm HESS} \approx 0.30^{\circ}$). 

\begin{figure*}
    \centering
    \includegraphics[width=0.49\linewidth]{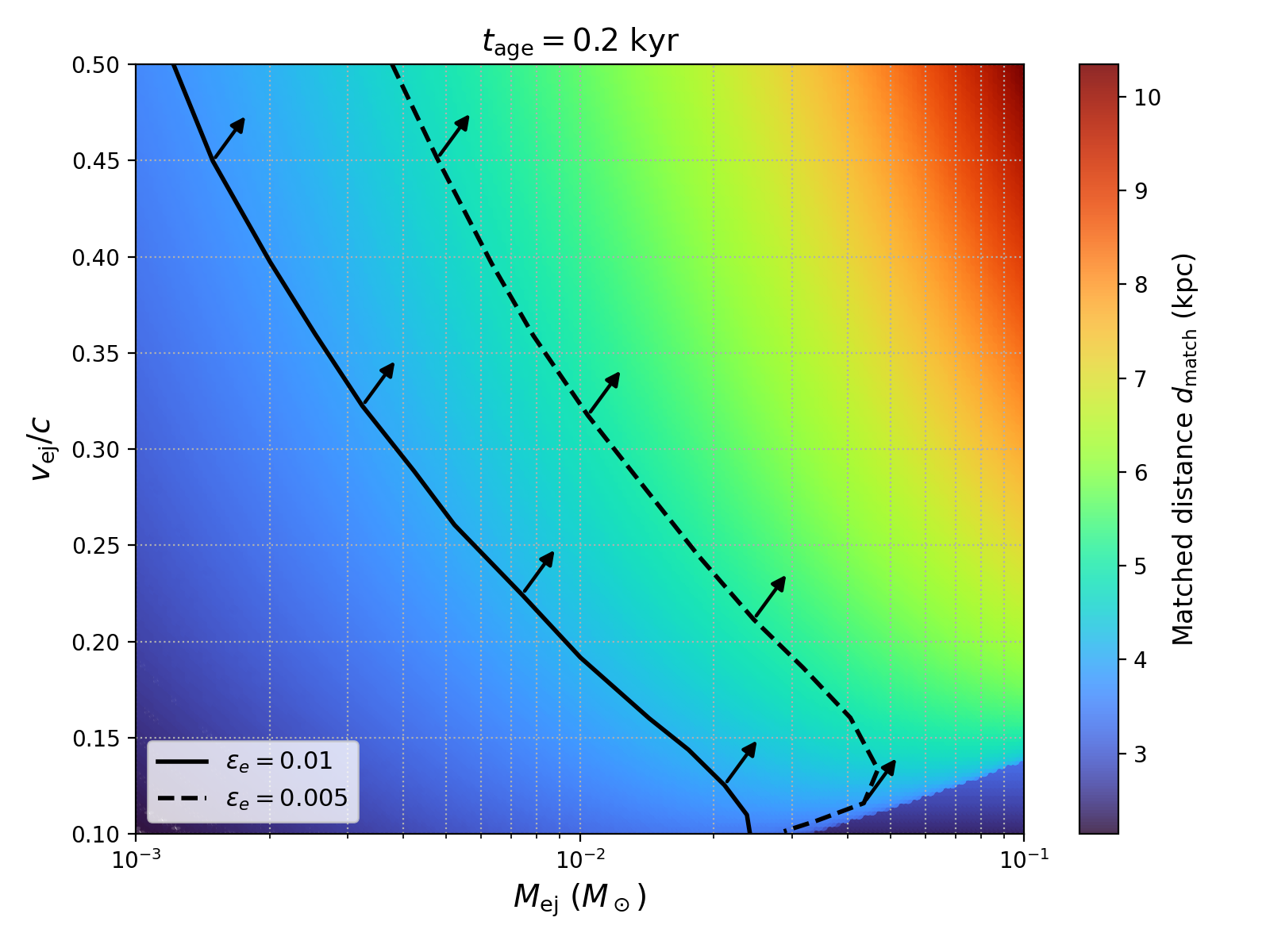}
    \includegraphics[width=0.49\linewidth]{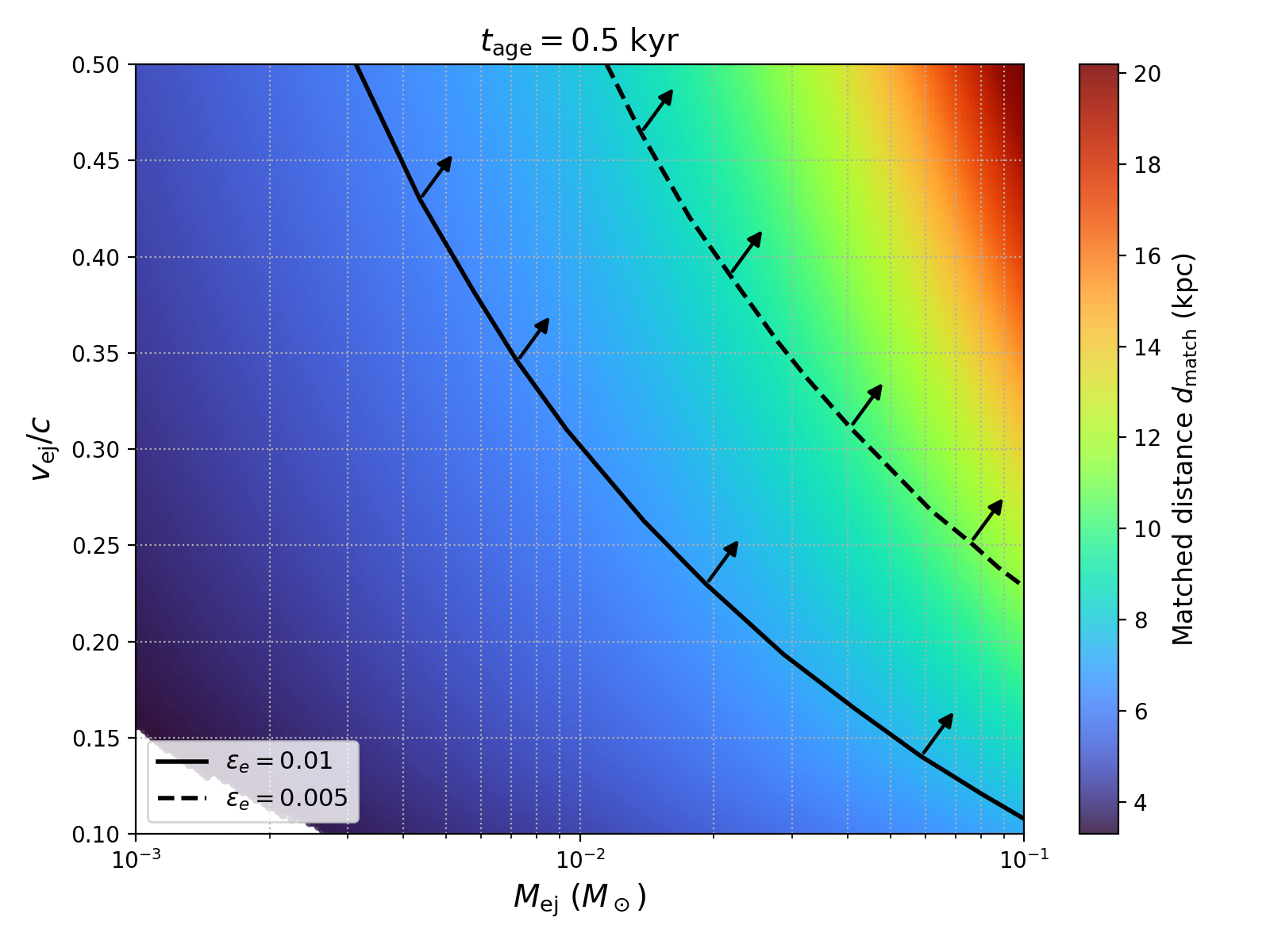}
    \includegraphics[width=0.49\linewidth]{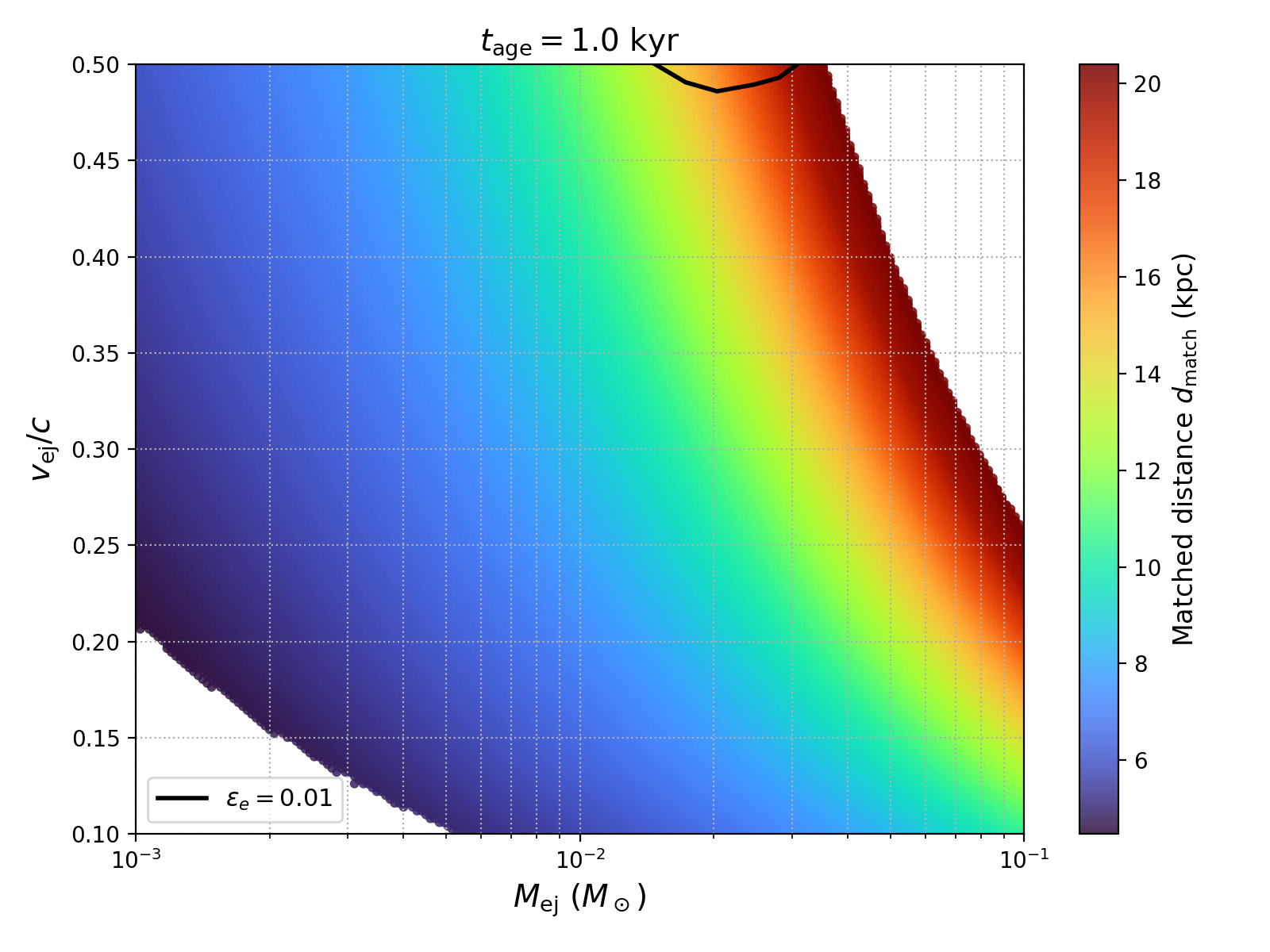}    
    \includegraphics[width=0.49\linewidth]{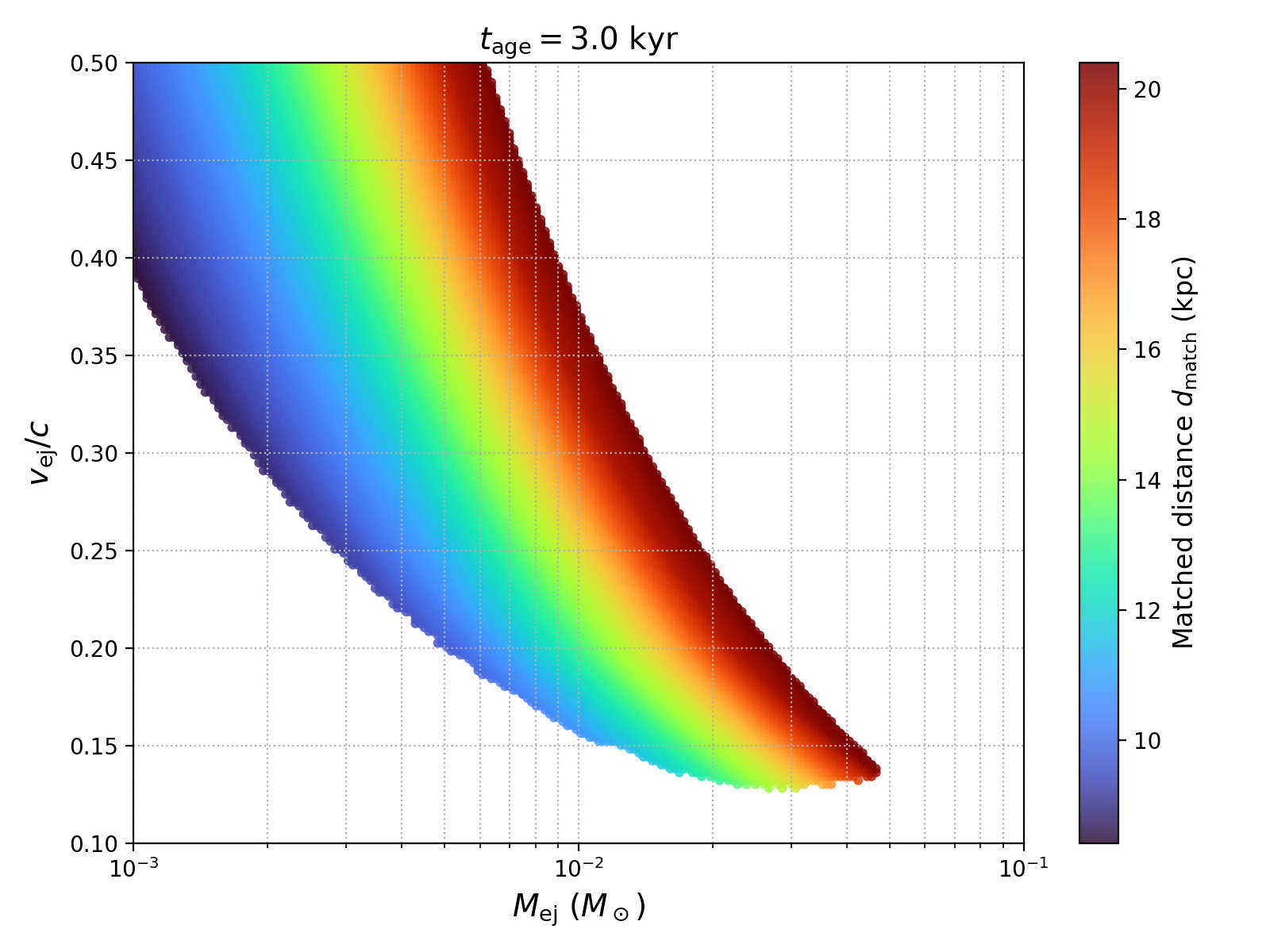}
    \caption{ Parameter space of $M_{\rm ej}-v_{\rm ej}$ corresponding to different KN age ($t_{\rm age}$) constrained with the observed angular size of HESS J1507-622 and a minimum ejecta energy cut off  $E_{\rm ej} \geq 2 \times 10^{50} ~(d/ 10~\mathrm{kpc})^2 ~ (\epsilon_{\rm e}/0.1)^{-1}$~erg. The color maps show the parameter spaces  for $\epsilon_{\rm e} =0.1$, while the allowed regions for  $\epsilon_{\rm e} =0.01$ and  $\epsilon_{\rm e} =0.005$ have also been shown by the black and black dashed curves with arrows, respectively. The parameter space shrinks rapidly with increasing $t_{\rm age}$ and hence a only tiny region survives for $t_{\rm age} = 1$~kyr, while completely disappears for $t_{\rm age} = 3$~kyr for $\epsilon_{\rm e} \le 0.01$. }
    \label{fig:parameter_space_with_energy_cutoff2}
\end{figure*}

The physical radius of the KNR is $r_{\rm KNR}=(d_{\rm HESS}/2)\tan\theta_{\rm HESS}$. Since the distance $d_{\rm HESS}$ is unknown, we treat it as a free parameter ranging from 1-20 kpc. We model the KNR radius evolution similarly to SNR~\citep{1988ApJ...334..252C,2019ApJ...880...23W} as 
\begin{equation}
    r_{\rm KNR}(t) = v_{\rm ej}t +\left[\dfrac{2.026E(t-t_{\rm ST})^2}{\rho} \right]^{1/5} \, ,
    \label{eq:radius}
\end{equation}
for the Sedov-Taylor (ST) phase whose  characteristic timescales is given by, 
\begin{equation}
t_{\rm ST}
=
69.5
\left(\frac{0.1c}{v_{\rm ej}}\right)
\left(\frac{M_{\rm ej}}{M_\odot}\right)^{1/3}
\left(\frac{n_{\rm ISM}}{1~{\rm cm}^{-3}}\right)^{-1/3}
{\rm yr} \, .
\label{eq:tsw}
\end{equation}
The local ISM density for the HESS source is approximated as $n_{\rm ISM}(d_{\rm HESS}) \approx 0.46 (392-70 d_{\rm HESS} + 6 d_{\rm HESS}^2)^{-1.065} \, \mathrm{cm}^{-3}$, where $d_{\rm HESS}$ is in kpc \citep{2013ApJ...770..118M}. For a given age $t = t_{\rm age}$ and distance $d_{\rm HESS}$, the derived $r_{\rm KNR}$ maps directly onto allowed values for $M_{\rm ej}$ and $v_{\rm ej}$ via Eqs.~[\ref{eq:radius}-\ref{eq:tsw}]. 

The TeV gamma-ray flux provides a second constraint via the electron energy fraction $\epsilon_{\rm e}\equiv E_{\rm tot}/E_{\rm ej}$, where $E_{\rm tot}$ is the total accelerated electron energy. We adopt a theoretical maximum of $10 \%$ for this fraction~\citep{Morlino:2021rzv,David:2022crk}.
For instance, a KNR with $E_{\rm tot} \approx 2 \times 10^{49}$~erg at $d_{\rm HESS} = 10$~kpc can successfully reproduce the observed gamma-ray flux using the model described in Sec.~\ref{subsec:CR_model}. This requires a minimal ejecta energy of $E_{\rm ej} = 2 \times 10^{50}$~erg. Since the observed flux scales as $E_{\rm ej}/d_{\rm HESS}^2$, $E_{\rm ej}$ scales as $d_{\rm HESS}^{2}$. Thus we define the following distance-dependent energy threshold: $E_{\rm ej} \gtrsim 2 \times 10^{50} (d/ 10~\mathrm{kpc})^2$~erg.  

Imposing both the angular size and energy constraints, we derive the allowed ($M_{\rm ej}-v_{\rm ej}$) parameter space for assumed KNR ages of $t_{\rm age}=0.2, 0.5, 1.0$ and $2.0 ~\rm kyr$ (Fig.~\ref{fig:parameter_space_with_energy_cutoff2}) with $\epsilon_{\rm e}=0.1$. The color bar indicates required distance $d_{\rm match}$ for each parameter combination. Generally, the allowed phase space shrinks significantly for older KNRs ($t_{\rm age}\gtrsim1$kyr): 
a larger $t_{\rm age}$ implies a smaller $V_{\rm ej}$ \citep{1982ApJ...258..790C} and a larger physical radius; while the former lowers the required ejecta energy, the latter requires a larger distance to match the fixed angular extent. Furthermore, because the actual electron energy fraction $\epsilon_{\rm e}$ might be lower than our $10\%$ upper limit \citep{Morlino:2021rzv,David:2022crk}, Fig.~\ref{fig:parameter_space_with_energy_cutoff2} also illustrates the allowed region for more conservative efficiencies of $\epsilon_{\rm e} =0.01$ and 0.005. Reducing $\epsilon_{\rm e}$ severely constricts the lower bound of the viable parameter space. This restriction increases rapidly with $t_{\rm age}$; for $\epsilon_{\rm e} \le 0.01$, only a tiny allowed region survives at $t_{\rm age} = 1$~kyr, and the viable parameter space disappears entirely for $t_{\rm age} \geq 3$~kyr.

\begin{table}[htbp]
\centering
\caption{Allowed distance range and most likely distance for HESS J1507$-$622 under the condition $E_{\rm ej}\geq 8\times10^{49}\,\rm erg$. }
\label{tab:distance_constraints}
\begin{tabular}{cccc}
\hline
\hline
Age (kyr) &
$d_{\rm min}$ (kpc) &
$d_{\rm ml}$ (kpc) &
$d_{\rm max}$ (kpc) \\
\hline
0.2 & 2.1 & 3.8 & 10.3 \\
0.3 & 2.6 & 5.0 & 13.8 \\
0.4 & 3.0 & 6.0 & 17.5 \\
0.5 & 3.3 & 6.4 & 20.0 \\
0.6 & 3.6 & 7.2 & 20.0 \\
0.7 & 3.8 & 7.4 & 20.0 \\
0.8 & 4.0 & 6.5 & 20.0 \\
0.9 & 4.3 & 7.3 & 20.0 \\
1.0 & 4.5 & 6.9 & 20.0 \\
2.0 & 6.3 & 9.0 & 20.0 \\
3.0 & 8.4 & 14.3 & 20.0 \\
\hline
\end{tabular}
\end{table}

Table~\ref{tab:distance_constraints} summarizes the minimum ($d_{\rm min}$), maximum ($d_{\rm max}$), and most likely ($d_{\rm ml}$) distances of the source with various assumed ages. We derive $d_{\rm ml}$ via  Gaussian kernel density estimation~\citep{Silverman1986}, while $d_{\rm max}$ is limited by  either the source's angular size or the maximum extent of stellar disk from the Galactic center ($20$~kpc), whichever is smaller. For ages $t_{\rm age} = (0.2-3.0)$~kyr, the most likely distances of HESS J1507$-$622 fall between $3.6$ and $14.5$~kpc if it is a KNR.

This analysis confirms that a KNR provides adequate energetics to accelerate electrons and reproduce the observed properties of HESS J1507$-$622. To further validate this scenario, we next estimate the detailed gamma-ray spectra for these most likely configurations using the framework outlined in Sec.~\ref{subsec:CR_model}.

\begin{table}[t]
\centering
\caption{Representative KNR model parameters used for gamma-ray flux calculation.}
\label{tab:KNR_parameters_CRs}
\setlength{\tabcolsep}{3.5pt}
\begin{tabular}{cccccc}
\hline\hline
$t_{\rm age}$ & $d_{\rm ml}$ & $E_{\rm tot}$ & $\epsilon_{\rm e}$ 
& $E_{\rm ej}$ & $E_{\rm e,max}$ \\
(kyr) & (kpc) & ($10^{48}$ erg) & 
& ($10^{50}$ erg) & (TeV) \\
\hline
0.2 & 3.8  & 2.7 & 0.01 & 2.7 & 622 \\
0.5 & 6.4  & 7.8 & 0.01 & 7.8 & 439 \\
1.0 & 6.9  & 9.2 & 0.05 & 1.8 & 218 \\
3.0 & 14.3 & 43  & 0.10 & 4.3 & 130 \\
\hline
\end{tabular}
\end{table}

\begin{figure*}
    \centering
    \includegraphics[width=0.49\linewidth]{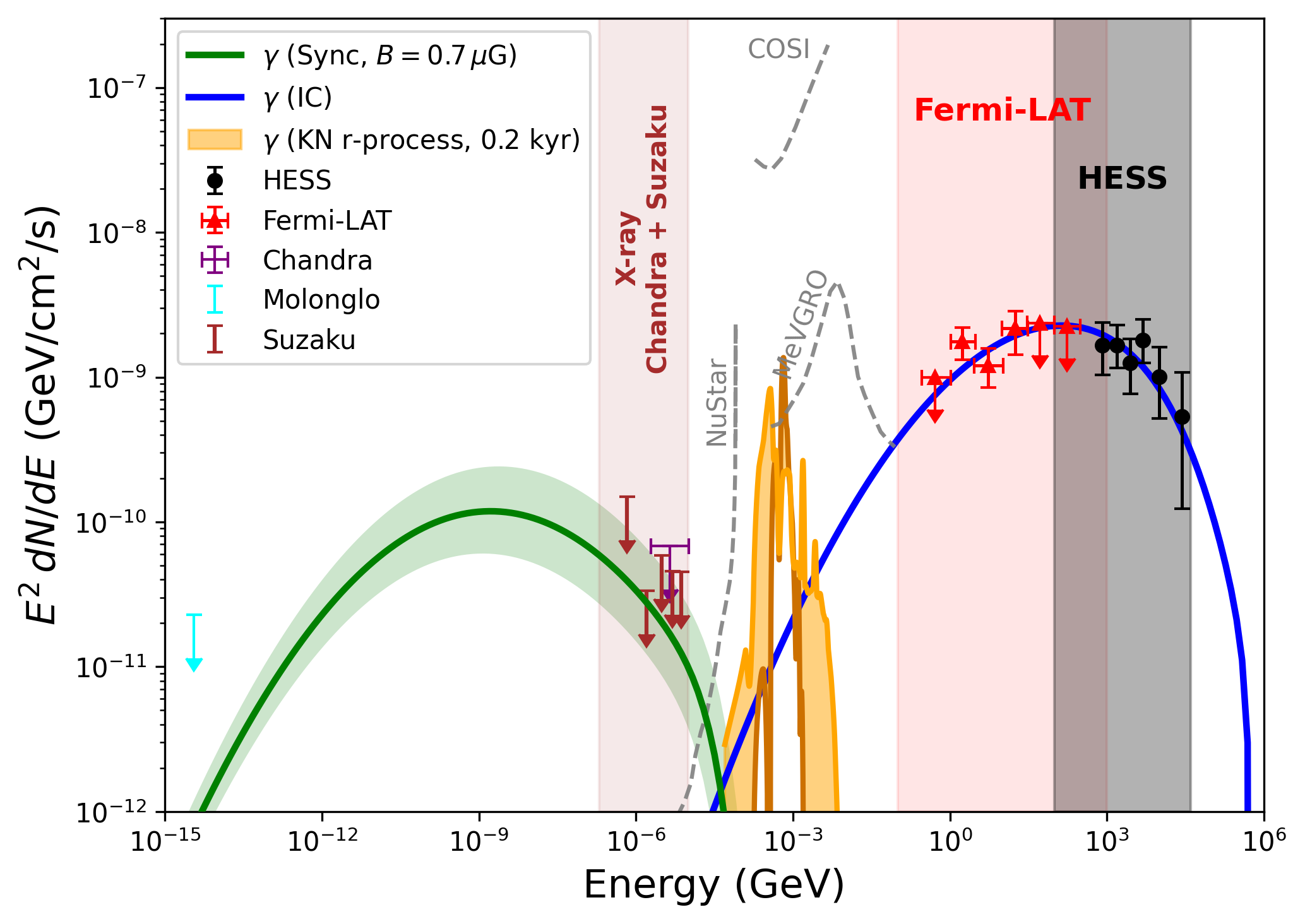}
    \includegraphics[width=0.49\linewidth]{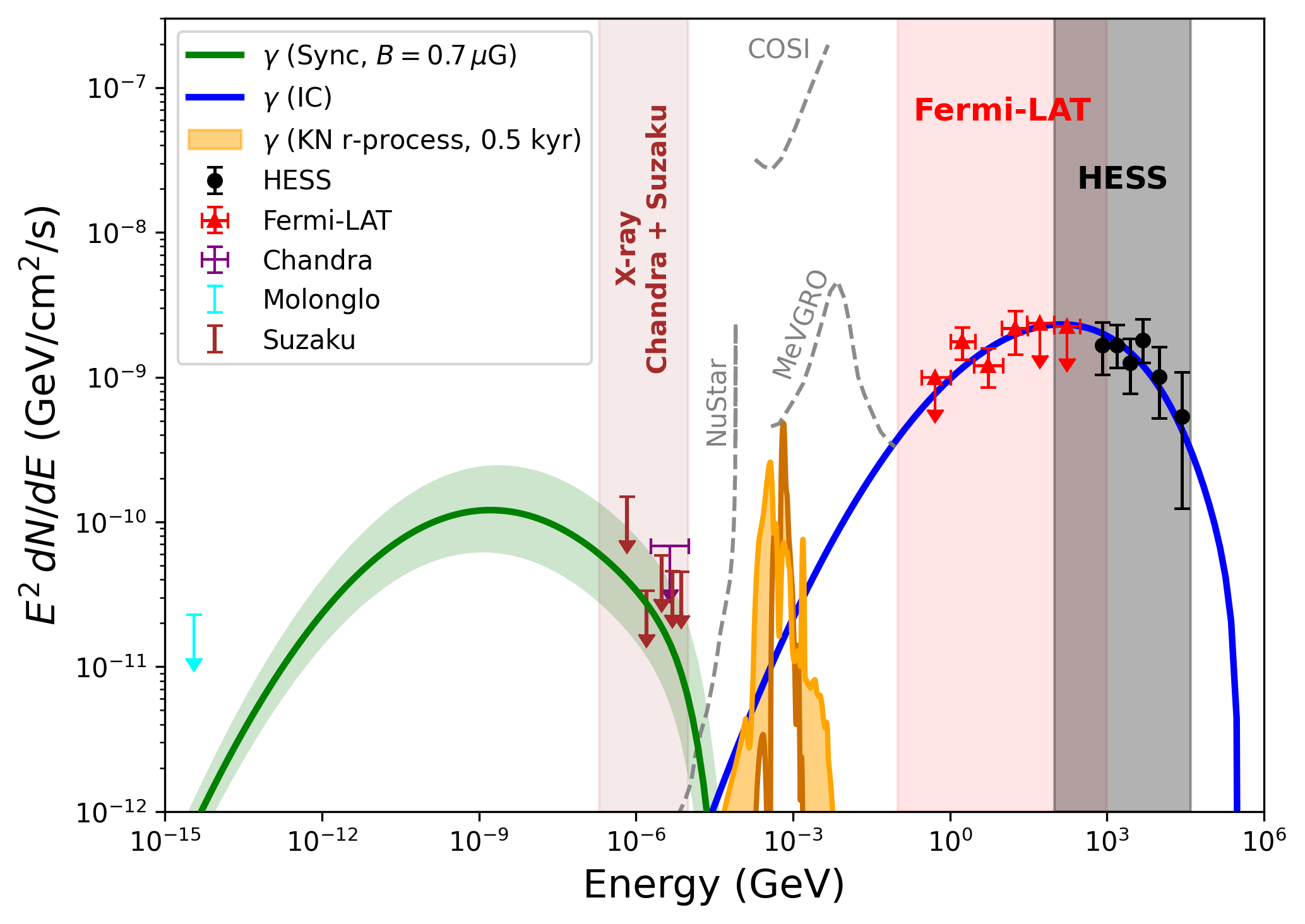}
    \includegraphics[width=0.49\linewidth]{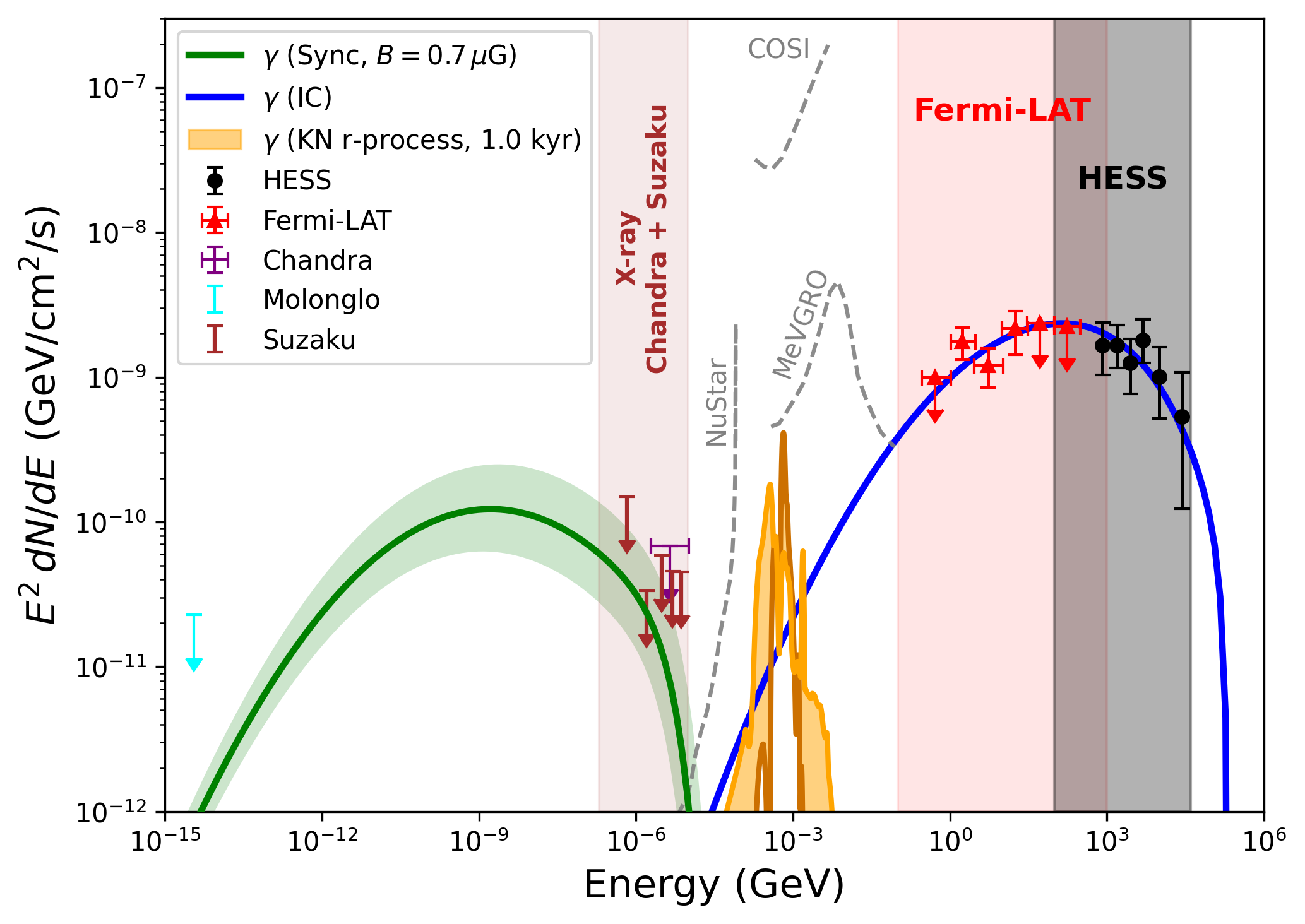}
    \includegraphics[width=0.49\linewidth]{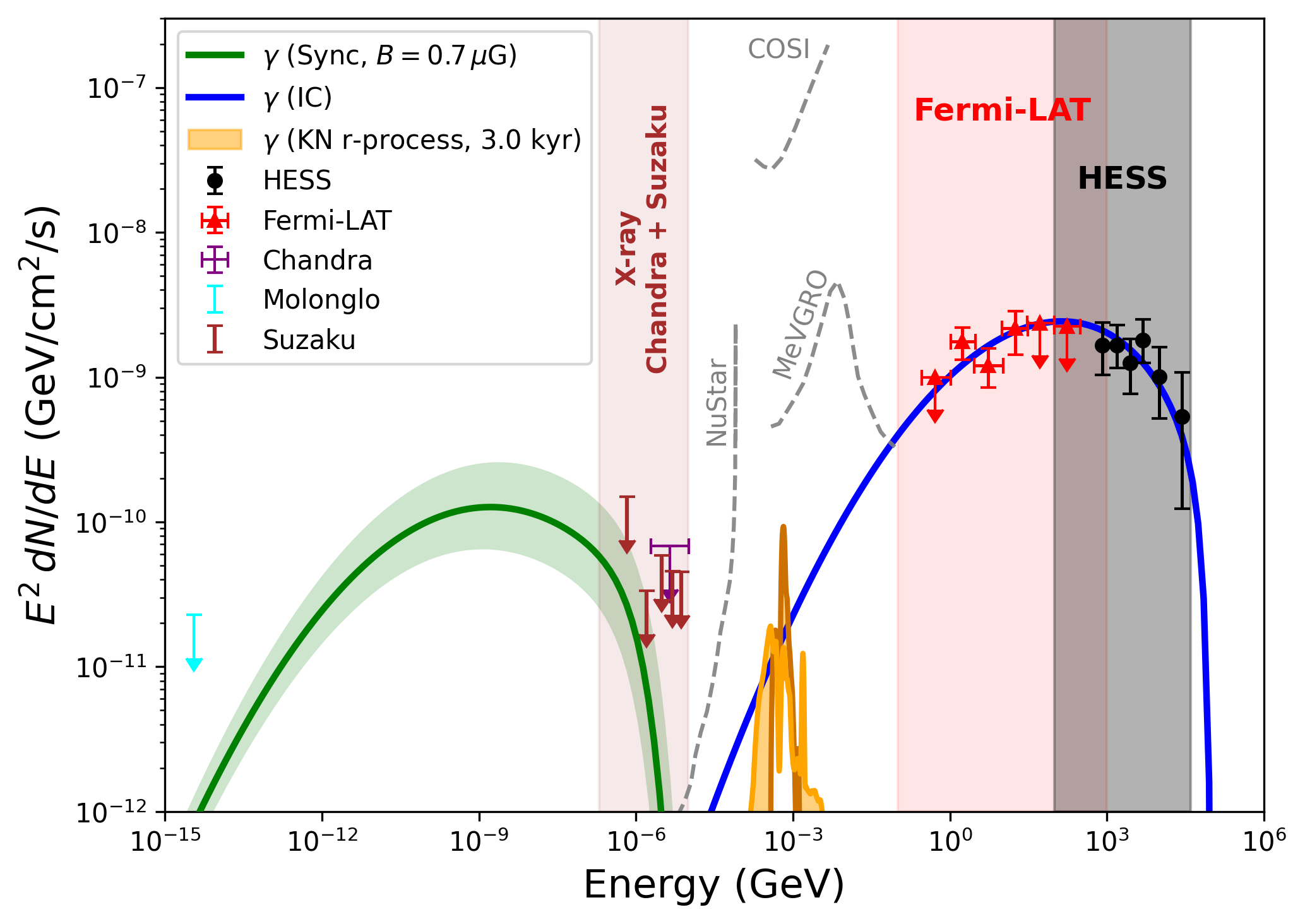}
    \caption{Gamma-ray flux from KNR produced via synchrotron (green;  band: $B= 0.5-1.0~\rm \mu G$) and inverse Compton (blue) radiation corresponding to different  KNR ages (\textit{Top left:} $0.2$~kyr, \textit{Top right:} $0.5$~kyr, \textit{Bottom left:} $1.0$~kyr, and \textit{Top left:} $3.0$~kyr), together with the gamma-ray observations of HESS J1507$-$622  by \textit{Fermi}-LAT (red data points) and HESS (black data points). We also show, as upper limits, the X-ray measurements of CXOU J150850.6$-$621018~\citep{2012A&A...545A..94D} and CXOU J150706.0$-$621443~\citep{Eger:2014eaa} obtained with Chandra (purple) and Suzaku (maroon), respectively, together with the radio upper limits from Molonglo~\citep{2012A&A...545A..94D}. 
    Additionally, we show the corresponding $r$-process gamma-ray emission in orange, with the band representing the associated  model uncertainties.
    The dark orange curve shows the flux from weak $r$-process production (corresponding to kilonova outflows with $Y_e = 0.3$ as described in \cite{Wang2020}), while light orange curve is for robust $r$-process production (corresponding to the outflows with $Y_e=0.15$). For future detection perspectives, we show the sensitivities of MeVGRO~\citep{MEVGRO}, NuStar~\citep{2005SPIE.5900..266K,Lucchetta:2022nrm}, and COSI~\citep{Tomsick:2023aue} by the gray dashed lines. }
    \label{fig:Point_sources}
\end{figure*}

\subsection{Results}

To model the broadband emission of HESS J1507$-$622, we select four representative KNR ages ($t_{\rm age}=0.2$, $0.5$, $1.0$, and $3.0$~kyr) at their most likely distances derived in Table~\ref{tab:distance_constraints}. Table~\ref{tab:KNR_parameters_CRs} lists a viable (not unique) set of physical parameters ($E_{\rm tot}$, $E_{\rm ej}$, and $E_{\rm e, max}$) that lie within the allowed phase space and successfully reproduce the observed gamma-ray flux. 
We assume a primary electron spectrum with indices $\alpha=2.5$ and $\beta=0.3$, consistent with standard log-parabolic models~\citep{Massaro:2005qg}. The resulting multi-wavelength spectra are presented in Figure~\ref{fig:Point_sources}, with the IC and synchrotron components shown as blue and green curves, respectively.

As shown in Figure~\ref{fig:Point_sources}, the IC emission provides a good fit to the GeV-TeV gamma-ray data from \textit{Fermi}-LAT (black points) and HESS (red points). While current observations have already provided precise measurements of the source's gamma-ray spectrum, the upcoming CTAO-South array\footnote{https://www.ctao.org/for-scientists/performance/} will enable a more detailed probe of its nature. For the synchrotron component, we adopt a fiducial ambient magnetic field $B=0.7~\mu{\rm G}$ and explore a range of $0.5$--$1~\mu{\rm G}$ to illustrate the dependence of resulting signal on $B$. The predicted synchrotron emission extends primarily across the radio, optical, and X-ray bands. Since no firmly established counterpart to HESS J1507$-$622 has been identified at these wavelengths, we conservatively treat available observations as upper limits. Our model safely respects these constraints, including the X-ray upper limits from Chandra~\citep[maroon arrows;][]{2012A&A...545A..94D} and Suzaku~\citep[purple arrow;][]{Eger:2014eaa}, as well as the radio upper limit from Molonglo~\citep[cyan arrows;][]{2012A&A...545A..94D}.

While our leptonic model successfully reproduces the gamma-ray observations and lower-energy constraint of HESS J1507$-$622 by HESS and \textit{Fermi}-LAT and alleviates the limitations of PWN and SNR interpretations, a definitive test of the KNR scenario relies on a unique multi-mesesenger signature: MeV gamma rays from rapid neutron capture ($r$-process) nucleosynthesis. 
The neutron-rich ejecta from a neutron star merger produces some thousandths to hundredths of a solar mass of highly radioactive, heavy nuclear species through the $r$-process. The signature kilonova optical emission is the result of $r$-process radioactive decay products thermalized in lanthanide-rich ejecta \citep{Kasen+2017}.  As the remnant expands and becomes optically thin, the $\sim$ MeV gamma rays emitted in nuclear $\alpha$ decays, $\beta$ decays, and fission can provide a unique and direct $r$-process signature \citep{Hotokezaka+2016,Li2019_MeV,Korobkin+2020,Wang2020,Chen2021_MeV, Terada2022_MeV, VasshTl208,Gross2025,Lariviere2026}. 
While GW170817 was too distant for nuclear gamma rays to be detected, a Galactic event or remnant may be observable in next generation detectors \citep{Korobkin+2020, Lariviere2026}.

In Fig.~\ref{fig:Point_sources} (orange curves), we present predicted $r$-process gamma-ray spectra corresponding to our selected KNR ages and distances. To bracket the uncertainties in outflow conditions/compositions, we show two scenarios: the dark orange curve shows the spectrum from weak $r$-process production (outflows with electron fraction $Y_e$
=0.3 as described in \cite{Wang2020}), while the light orange curve represents robust $r$-process production ($Y_e$
=0.15). At energies near 1 MeV, the weak $r$-process yields a higher gamma-ray flux than the robust case. In contrast, the robust $r$-process produces relatively stronger signals at both lower and higher energies, although these emissions are much fainter than the peak around 1 MeV. Because the $r$-process nuclei decay over time, the signal is stronger for younger KNRs; remnants younger than $\sim1$ kyr could  potentially be detected by the next generation MeV observatories such as MeVGRO \citep{MEVGRO}. However, the predicted fluxes remain well below the sensitivity limit of the upcoming COSI \citep{Tomsick:2023aue} and are also too faint for NuSTAR observation energy range \citep{2005SPIE.5900..266K,Lucchetta:2022nrm}.

\section{Detectability of the Kilonova and Kilonova Remnant}
\label{sec:KN_detectability}

Having demonstrated that HESS J1507$-$622 is a plausible KNR candidate, for which the future detection of MeV gamma rays would serve as an ultimate smoking gun, two natural questions arise: Are there any historical records of the original merger event? And what other future observations could help confirm its KN origin? In this section, we address why such a relatively young KN might have escaped historical detection and analyze the detectability of complementary KN dust signatures (thermal emission and light echoes).

\subsection{Historical kilonova detectability}

To estimate the expected brightness of the progenitor KN, we use GRB 170817A/GW170817 as a reference: this transient occurred at a distance of $40$~Mpc and reached its maximum brightness approximately one day after the merger, with a peak V-band apparent magnitude of $\sim 17$~\citep{Arcavi:2017xiz}. Rescaling its peak optical magnitude and afterglow emissions~\citep[detected in both optical and radio bands at around 150 days with optical apparent magnitude $\sim 26$, and the radio peak flux of $\sim 0.1~\rm mJy$ at frequencies of $3$--$6$~GHz;][]{Lamb:2018qfn} to our inferred KNR distance range of 3.8-14.3 kpc, we estimate a peak optical apparent magnitude between $\sim -0.8$ and $-3.2$. The late-time optical afterglow would reach
$5.7$--$8.2$, and the radio afterglow flux would peak at $\sim10^{3}$--$10^{4}$~Jy. Note that these estimates assume that the putative KN has intrinsic optical and radio properties comparable to those of GRB 170817A/GW170817 and neglect line-of-sight extinction and other environmental effects.

Had such a KN occurred in the Milky Way within the past $200$--$3000$ years, it could have appeared in the night sky with a brightness comparable to that of the brightest planets. For comparison, the maximum apparent magnitudes of Venus, Mars, and Jupiter are approximately $-4.9$, $-3.0$, and $-2.9$, respectively. Therefore, such a historical KN would, in principle, have been readily visible to the naked eye. Nevertheless, no obvious counterpart to such an event can be identified in the available historical astronomical records. Several factors could potentially account for the absence of such a record. First, KNe evolve rapidly in optical, fading at a rate of $\sim$1.1 mag/day as inferred from GRB 170817A/GW170817 \citep{Valenti+2017}, making them easy to miss during poor weather or bright lunar phases. 

Second, the source's  far southern declination ($\mathrm{Dec.}\simeq-62^{\circ}$) was inaccessible, or observable only at very low elevations, to Nothern Hemisphere cultures with well-documented historical astronomical records, including China, Japan, Egypt, and Europe \citep{Clark+1977}. 
While Southern Hemisphere cultures (including Indigenous Australian, M\=aori, San, and Andean peoples) held rich astronomical traditions, their sky lore prioritized seasonal markers, navigational stars, and structural `dark cloud' constellations \citep{Urton1981,Lewis-Williams2002,Harris+2013}. A static point of light appearing and vanishing in under 9 days leaves no orbital motion (unlike a comet) and fades too quickly to become integrated into generational oral history or symbolic rock art \citep{Hamacher2014}.
This observational bias is also reflected in the historical record of Galactic SNe. Among the well-established historical events, SN~1006, at $\mathrm{Dec.}\simeq-42^{\circ}$, is the southernmost example. Thus, the lack of systematic astronomical records from the Southern Hemisphere further reduces the likelihood that a transient at the position of HESS J1507$-$622 would have been documented.

More generally, the absence of a historical record does not preclude a recent Galactic stellar explosion. Many young SNRs have no historical counterparts; interestingly, the youngest known Glactic SNR, G1.9+0.3~\citep[$\sim100$--$150$~yr;  $d\sim8.5$~kpc;][]{Reynolds:2008if}, escaped detection despite an expected unextinguished peak magnitude of $\sim-4.7$, likely due to severe line-of-sight extinction toward the Galactic center. Combined with the rapid fading of the transient phase and incomplete historical sky coverage in the far south, unfavorable observing conditions, and potential optical extinction, these factors mutually explain the missing historical record. It therefore remains entirely plausible that HESS J1507-622 originated from a KN event within the past $\sim200$--$3000$ years.

\subsection{Kilonova remnant detectability}

While the initial KN transient faded long ago, its expanding remnant can produce observable multi-wavelength signatures. Beyond the GeV-TeV gamma-ray emission, thermal dust emission and a light echo could provide complementary probes of the KNR nature.

\subsubsection{Thermal emission from dust}
 
Low-energy thermal emission in the radio, infrared, and optical bands, arising from processes such as free–free emission, thermal dust emission, and thermal bremsstrahlung, is strongly dependent on the density of the ambient medium. Given the low ISM density surrounding HESS J1507$-$622, such emission is likely extremely faint and below current detection limits. 

Because Galactic KNR dust properties remain unknown, we adopt two off-plane SNRs of comparable age as benchmark scenarios: Cas A ($\sim400$yr;$d=3.4$~kpc; off-plane distance $z\sim100$~pc) and SN 1006 ($\sim1000$yr;$d=2.2$~kpc; $z\sim500$~pc). We primarily focus on the mid-IR band, where extinction is negligible, as optical emission is heavily attenuated by dust and radio bands are typically dominated by non-thermal processes \citep{Dubner:2015aqa}. 
The $24~\mu \rm m$ surface brightness of SN 1006 measured by the Spitzer space telescope~\citep{2013ApJ...764..156W} is $(0.25-0.3)~\rm MJy~sr^{-1}$, whereas Cas A exhibits a much brighter $25.5 ~\mu \rm m$ emission of $\sim (10^3-10^4)~\rm  MJy~sr^{-1}$ as measured by JWST MIRI. This large difference likely arises from the warmer, denser dust in the younger Cas A remnant. Scaling these values to our inferred KNR distance range of $3.8-14.3$~kpc yields an expected mid-IR surface brightness of $\sim (10^{-2}-10^{4}) ~\rm MJy~sr^{-1}$ for HESS J1507$-$622. 

\subsubsection{Light echo}
Another possible observational signature of a KNR is a light echo, which is the radiation from the original transient scattered by surrounding interstellar dust and arriving at Earth with a delay relative to the prompt emission from the transient. This phenomenon has been observed in several SNRs, including SN 1987A \citep{1987ASparks,1987ABond}. A potential light echo from HESS J1507$-$622 would provide a delayed observable signature that could help constrain the KNR age and surrounding environment.

Following~\cite{Patat2005}, the echo magnitude ($m_{\rm echo}$) relates to the peak optical magnitude $m_{\rm peak}$ as:
\begin{equation}
    m_{\rm echo} \approx m_{\rm peak} + 2.5 \log_{10} \left (  \frac{R}{3 \tau_{\rm d} \Delta t_{\rm KN}  c}  \right)  \ ,
\end{equation}
where, $R$, $\tau_{\rm d}$, and $\Delta t_{\rm KN}$ are the distance to the scattering dust from the source, dust optical depth, and the KN duration. Because $m_{\rm echo}$ represents the integrated light over an extended scattering arc, 
the surface brightness ($\mu_{\rm echo}$) provides a more appropriate metric for detectability:
\begin{equation}
    \mu_{\rm echo} = m_{\rm echo} + 2.5 \log_{10} \left[ \frac{A_{\rm echo}}{1 ~\mathrm{arcsec}^2} \right] \ ,
\end{equation}
where, $A_{\rm echo}$ is the area of the scattering arc. 

To estimate $\mu_{\rm echo}$ for HESS J1507-622, we adopt representative values from GRB 170817 ($m_{\rm peak}=1$--$4$ and $\Delta t_{\rm KN}\approx2$ days) and the well-studied light echo of SN 1987A ($\tau_{\rm d}=3\times10^{-4}$; $A_{\rm echo}\approx100$--$1000\ {\rm arcsec}^2$). Assuming the characteristic dust radius $R$ reflects the KNR age and ranges from $0.2-3$ ${\rm kly}$, we predict a surface brightness of $\mu_{\rm echo}\approx27-33~\rm mag~arcsec^{-2}$.

Detecting such a light echo associated with HESS J1507$-$622 is highly feasible with modern deep-imaging facilities that can reach surface brightness limits of $\mu_{\rm echo}\lesssim30~\rm mag~arcsec^{-2}$~\cite{2017ApJ...834...16M}.
For this southern source, optical facilities in the Southern Hemisphere such as VLT/FORS2, Gemini South/GMOS-S, Magellan/IMACS, and Blanco/DECam are well-suited for targeted searches. Furthermore, the Vera C. Rubin Observatory/LSSTCam is expected to reach $\sim (30-31)~{\rm mag~arcsec^{-2}}$ on \(\sim10''\) angular scales in deep coadded observations~\citep{Yoachim2022SurfaceBrightness,2022MNRAS.513.1459M}.  
With dedicated background-control strategies, large-aperture telescopes have demonstrated sensitivities beyond $31~{\rm mag\,arcsec^{-2}}$, with surface-brightness profiles traced to $\sim33~{\rm mag~arcsec^{-2}}$~\citep{Trujillo:2015nna,Watkins_2024,laine2018lsstcadenceoptimizationwhite}. 
Space-based platforms are equally promising: the Hubble Space Telescope can reach $\sim 29~{\rm mag\,arcsec^{-2}}$ \citep{STScI_ACS_F606W}, 
and the recently launched Nancy Grace Roman Space Telescope\footnote{https://science.nasa.gov/mission/roman-space-telescope/} will offer comparable sensitivity with the bluest filter.

Ultimately, a deep targeted search for this light echo would provide a critical test of the KNR hypothesis. If detected, the echo would yield insights far beyond a simple time delay: it could reveal the early-time optical spectrum of the KN event \citep{Shappee:2017zly}. Identifying the spectroscopic signature of $r$-process nuclei within this echo would definitely confirm the source's KN origin. In contrast, a rigorous non-detection at these depths would strictly constrain the scenario, implying either an older KNR (age $\gtrsim500$ yr) or a locally dust-poor environment.

\subsection{Multi-messenger prospects}
As discussed in Sec.~\ref{sec:KN_Model}, the extremely large $pp$ interaction length ($\sim 11$~Gpc) strongly disfavors a hadronic origin for gamma-ray flux in a low-density ISM. Consequently, the associated high-energy neutrino flux from $pp$ interactions is expected to be negligible. However, if an undiscovered dense molecular cloud exists along the line of sight~\citep{Sarmah:2023pld}, interactions of  KNR-accelerated CRs with such could simultaneously produce the observed gamma rays and a potentially detectable neutrino flux.

Using the neutrino--gamma-ray scaling relation given in Eq.~(2) of \citet{Ahlers:2013xia}, the peak gamma-ray flux is $\approx 2\times10^{-9}~{\rm GeV,cm^{-2},s^{-1}}$, predicting a single-flavour $\sim1$TeV neutrino flux of $\approx 10^{-9}~{\rm GeV,cm^{-2},s^{-1}}$. While challenging for IceCube due to the southern sky location, this flux may be within the reach of KM3NeT~\citep{KM3NeT:2024uhg} and the upcoming HUNT neutrino telescope~\citep{2026NIMPA108671374C} with a long exposure time of $\sim10$~yr.  
Conversely, a firm neutrino non-detection would place a strong limit on the hadronic contribution, validating the predominantly leptonic origin of the observed gamma-ray emission. 

Finally, the predicted neutrino flux also depends on the composition and spectrum of the CR population~\citep{Kimura:2018ggg,Komiya:2017uds}. SuperTIGER measurements show an enhancement in Galactic CR abundances for elements with $42\leq Z\leq54$ compared to standard source model~\citep{2022AdSpR..70.2666W}. Because several of these elements receive substantial contributions from the $r$-process~\citep{OSBORN20268337}, this excess could potentially be linked to Galactic $r$-process sources like the KNR discussed here. However, as KNR CR properties remain poorly constrained, exploring this impact in detail is beyond the scope of this paper.

\section{Summary and Conclusion}
\label{sec:conclusion}

HESS J1507$-$622 is an intriguing unidentified Galactic VHE source with off-plane location ($b = -3.47^{\circ}$). Its compactness ($\theta_{\rm HESS} \approx 0.30^{\circ}$), high-latitude location, and the absence of low-energy (radio and X-ray) counterparts challenge the conventional PWN or SNR interpretations. In this Letter, we have proposed that HESS J1507$-$622 is a relatively young Galactic KNR, with gamma-rays emission from leptonic processes and with its off-plane location naturally explained by the natal kick imparted to the progenitor BNS system.

By exploring the allowed KN ejecta mass and velocity ($M_{\rm ej}-v_{\rm ej}$) parameter space, we demonstrated that a leptonic KNR scenario can successfully reproduce the observed properties. Shock-accelerated electrons undergoing IC scattering match the GeV-TeV gamma-ray flux measured by \textit{Fermi}-LAT and HESS, while the corresponding synchrotron emission remains below current low-energy (radio and X-ray) upper limits. Our physical constraints suggest the KNR is $200-3000$ years old and located at a distance of $3.6-14.3$ kpc. Most importantly, our analysis favours a KNR younger than 1 kyr.
While we found no historical record of any such event with naked-eye observations, it could have been easily missed due to rapid fading of KN event, the source's far-southern declination, and potential line-of-sight extinction. 

Because the leptonic gamma-ray spectra of KNRs, PWNe, and SNRs can appear similar, confirming the KNR nature of HESS J1507$-$622 requires independent and unique kilonova-associated signatures. We have identified several promising multi-wavelength and multi-messenger avenues for future targeted observations:

\textit{MeV Gamma Rays.---} The decay of $r$-process radioisotopes in the KNR ejecta produces a distinct $\sim$1 MeV gamma-ray signature. While too faint for current NuSTAR~\citep{2005SPIE.5900..266K} and the upcoming COSI~\citep{Tomsick:2023aue}, this flux would be within the reach of next-generation MeV observatories like MeVGRO \citep{MEVGRO}.

\textit{Light Echoes.---} Delayed scattering of the initial KN light by interstellar dust could produce a detectable light echo with an estimated surface brightness of $\mu_{\rm echo}\approx27-33~\rm mag~arcsec^{-2}$. Facilities such as the Rubin Observatory ~\citep{Yoachim2022SurfaceBrightness,2022MNRAS.513.1459M}, Hubble \citep{STScI_ACS_F606W}, and the Roman space telescope\footnote{https://science.nasa.gov/mission/roman-space-telescope/} can reach these limits. Crucially, this echo would allow us to observe the early-time optical spectrum of the historical transient, offering a confirmation of $r$-process synthesized in the source.

\textit{Thermal Dust and Multi-messengers.---} Deep mid-IR observations could reveal thermal emission from the KNR dust, while the remnant itself may be a local heavy cosmic-ray accelerator. The latter could potentially explain the abundance enhancement of heavy ($42\leq Z\leq54$) Galactic cosmic rays observed by SuperTIGER~\citep{2022AdSpR..70.2666W}. Furthermore, if a dense molecular cloud exists along the line of sight, hadronic interactions could yield a steady TeV neutrino flux detectable by KM3NeT~\citep{KM3NeT:2024uhg} and the upcoming HUNT~\citep{2026NIMPA108671374C}.

Ultimately, a confirmed Galactic KNR would represent an unprecedented opportunity to study $r$-process nucleosynthesis, binary merger environments, and heavy particle acceleration in our own Galaxy. The ability of the KNR scenario to consistently account for the observed properties of HESS J1507$-$622 makes it a compelling hypothesis, underscoring the critical need for advanced MeV gamma-ray missions and deep optical/IR imaging to finally unveil the nature of this enigmatic source.

\begin{acknowledgments}
The authors would like to thank Brian Fields for the helpful and constructive suggestions. The work of P.S. and X.W. is supported by National Natural Science Foundation of China (Grant Nos. 12494570, 12494574 and 1252100) and the National Key R$\&$D Program of China (2021YFA0718500). The work of M-Y.G. is supported by National Natural Science Foundation of China (Grant Nos. 12673057). X.W., S-X.Y., and M-Y.G. also acknowledge support from the China's Space Origins Exploration Program. R.S. acknowledges support from the U.S. Department of Energy under Grant Nos. DE-FG02-95-ER40934 and DE-SC00268442 (ENAF) and the U.S. National Science Foundation under Grant Nos. PHY-2020275 (N3AS) and 21-16686 (NP3M).
\end{acknowledgments}

\bibliography{biblio}{}
\bibliographystyle{aasjournalv7}

\clearpage

\end{document}